# Carrier Lifetimes in a III-V-N Intermediate Band Semiconductor

J. N. Heyman[1], A. M. Schwartzberg[2], K. M. Yu[2,3], A.V. Luce[2,4], O. D. Dubon[2,4], Y. J. Kuang[5], C. W. Tu[5], W. Walukiewicz[2]

[1.] *Physics Department, Macalester College, St. Paul, MN 55105, USA.   Email: heyman@macalester.edu.*

[2.] *Materials Sciences Division, Lawrence Berkeley National Laboratory, Berkeley, CA, 94720, USA.*

[3.] *Department of Physics and Materials Science, City University of Hong Kong, Hong Kong SAR, China.*

[4.] *Department of Materials Science and Engineering, University of California, Berkeley, California 94720, USA.*

[5.] *Department of Physics, University of California, San Diego CA 92093, USA.*

## Abstract

We have used transient absorption spectroscopy to measure carrier lifetimes in the multiband band semiconductor $GaP_yAs_{1-x-y}N_x$. These measurements probe the electron populations in the conduction band, intermediate band and valance band as a function of time after an excitation pulse.   Following photoexcitation of $GaP_{0.32}As_{0.67}N_{0.01}$ we find that the electron population in the conduction band decays exponentially with a time constant $\tau_{CB}$ =23ps.  The electron population in the intermediate band exhibits bimolecular recombination with recombination constant $r$ = $2 \cdot 10^{-8}$ cm$^{-3}$/s.  In our experiment an optical pump pulse excited electrons from the valance band to the intermediate and conduction bands, and the change in interband absorption due to absorption saturation and induced absorption was probed with a delayed white light pulse.  We modeled the optical properties of our samples using the band anti-crossing model to extract carrier densities as a function of time.  These results indicate that the minority carrier lifetimes are too short for efficient solar power conversion and that improvements in material quality will be required for practical applications of $GaP_yAs_{1-x-y}N_x$ based intermediate band solar cells.

## 1.    Introduction

Dilute III-V Nitrides (III-V$_{1-x}$ N$_x$ with $x$ less than ~5%) are termed highly mismatched alloys because the size and electronegativity of the *N* atoms differ strongly from those of the group V host atoms.  These materials have been successfully described by the band anticrossing model (BAC) [1] [2], which predicts that the interaction between the extended conduction band states of the host material and the localized energy states of the isoelectronic nitrogen impurities splits the conduction band into a distinct upper band ($E_+$) and lower band ($E_-$) separated by a gap. For example, studies of GaAsN have shown [2, 3, 4, 5] that the band anticrossing model can explain the strong dependence of the bandgap and electron effective mass on nitrogen doping. Photomodulated reflection measurements have been used to identify the minima of the $E_+$ and $E_-$ bands in GaAsN and show that they shift with nitrogen concentration and hydrostatic pressure as predicted by the BAC model. Other III-V dilute nitrides including[1, 2, 6, 7] GaInAsN, GaPN, GaAsSbN and GaPAsN have been investigated and successfully described using the BAC model.

Dilute III-V nitrides are promising materials for intermediate band solar cells (IBSC). In the intermediate band concept[8-10] solar photons excite electrons to the conduction band through two processes: direct excitation from the valance band to the conduction band ($VB-E_+$), or stepwise excitation from the valance band to the intermediate band ($VB-E_-$), and then to the conduction band ($E_- - E_+$). By adjusting the material composition these three transitions can be matched to the solar spectrum to efficiently capture solar energy. In an IBSC the intermediate band material is inserted into a $p$-$n$ junction with bandgap matching the $VB-E_+$ gap so that photoelectrons and holes are extracted from the conduction band ($E_+$) and valance band. In principle, an intermediate band solar cell possesses the advantages of a three-junction solar cell in simpler device architecture. While the Shockley-Queisser efficiency limit for a single junction photovoltaic device in concentrated sunlight is 41%, the limit for an intermediate band solar cell is predicted[8] to be 63%. Intermediate band solar cells have been implemented using GaAsN [11] and ZnTeO [12] highly mismatched alloys and by using quantum dot arrays inserted into a III-V host material[13]. The status of the IBSC field has been recently reviewed by Y. Okada et. al.[10]

Recently, Lopez, *et. al.*, [11] have reported an IBSC based on GaAs$_{0.976}$ N$_{0.024}$. Photocurrent spectra clearly show response associated with the $VB-E_-$ and $VB- E_+$ transitions. Photocurrent generation by sequential excitation ($VB-E_-$ followed by $E_- - E_+$) has also been observed[14] in GaAsN, and the $E_+$- $E_-$ transition has been observed in electroluminescence.[15] The efficiency of Lopez's prototype device was relatively low, and the open circuit output voltage (0.9V) was below prediction (~1.6V). The authors suggested that the low output voltage likely results from short minority carrier lifetimes. In GaAsN the BAC theory predicts formation of a wide intermediate band with a finite density of states between the $E_-$ and $E_+$. These gap states may be a pathway for rapid electron relaxation from the conduction band to the intermediate band. We also note that the energy band positions in GaAsN are not well matched to the solar spectrum.

In ZnTeO the addition of oxygen at low concentrations produces a narrow band of oxygen-derived extended states below the conduction band edge of ZnTe. This intermediate band can be described well by the BAC model. [16] Tanaka, *et. al.*, [12 17] have demonstrated a ZnTeO IBSC. They used photomodulated reflectance measurements to determine the critical energies of the $VB-E_-$, $VB–E_+$ and $E_- - E_+$ transitions. Photocurrent spectra of their device showed distinct photocurrent generation from $VB-E_-$ and $VB-E_+$ transitions. The external quantum efficiency was low, and the open circuit output voltage (0.36V) was much lower than predicted ($\sim$2V). While the demonstration of these prototype IBSC devices is an important proof of principle, further development is required to make the IBSC concept viable. Significant improvements may result from optimizing absorption in the intermediate band material to match the solar spectrum, and by improving carrier lifetimes.

GaPAsN quaternary alloy is a promising material for implementing the IBSC concept. [18] Material with P alloy fraction $\sim$40% and N fraction $\sim$2% is predicted to provide a nearly optimum band structure for solar energy production. The nitrogen defect level in this material lies near or below the conduction band minimum in the host material, so that the anticrossing interaction produces a narrow N-derived intermediate band that is well separated from conduction band. Kuang, *et. al*., [19] have recently grown GaPAsN thin films on GaP substates with gas-source molecular beam epitaxy. Photomodulated reflectance (PR) measurements found $VB$-$E_-$ and $VB$-$E_+$ transitions close to transition energies predicted by the BAC model. [7] Optical transmission measurements showed strong $VB$-$E_-$ absorption. However, $VB$-$E_+$ transitions were not observed in transmission because the GaP substrates strongly absorb in this spectral region. Also, intermediate band to conduction band transitions ($E_-$ - $E_+$) were not observed in absorption or PR because these require population of the intermediate band.

Carrier dynamics in GaPAsN have recently been investigated by Baranowski, *et. al*. using time-resolved photoluminescence [20]. They report luminescence near the $VB$-$E_-$ absorption edge and measure luminescence lifetimes of $\sim$100ps at room temperature. The lifetimes increase dramatically as the temperature is reduced, reaching $\sim$50ns at T <100K. The authors describe the temperature dependent luminescence using a hopping exciton model: Excitons form after photoexcitation and are rapidly localized. Thermally activated hopping moves excitons between shallow localized states and deep non-radiative recombination centers. At high temperatures, all trap sites are accessible, and the deep non-radiative recombination sites dominate recombination. At low temperatures excitons trapped in shallow localized states are protected from the deep traps and long luminescence lifetimes are observed. This work did not determine the conduction band carrier lifetime because luminescence initiating in the conduction band was not observed. Also, luminescence primarily probes the dynamics of excitons rather than free carriers and can be dominated by defect-bound excitons.

In this paper, we use time-resolved optical absorption to investigate free carrier dynamics in two GaP$_y$As$_{1-x-y}$N$_x$ samples with compositions close to optimum for solar energy collection. Samples were selected having unperturbed conduction bands slightly below (S205B) or slightly above (S376B) the N-defect level. Transient absorption primarily probes carriers in the bands and can determine carrier populations and distribution functions. We observe saturation of the $VB$-$E_-$ and $VB$-$E_+$ transitions, as well as induced absorption at the $E_-$-$E_+$ transition energy due to optical population of the intermediate band. This allows us to find the carrier populations following photoexcitation and determine photocarrier lifetimes in the conduction and intermediate bands.

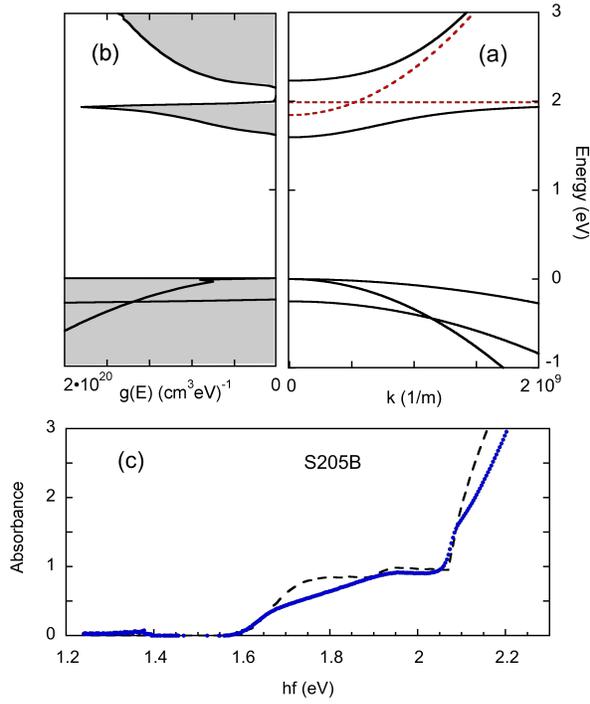

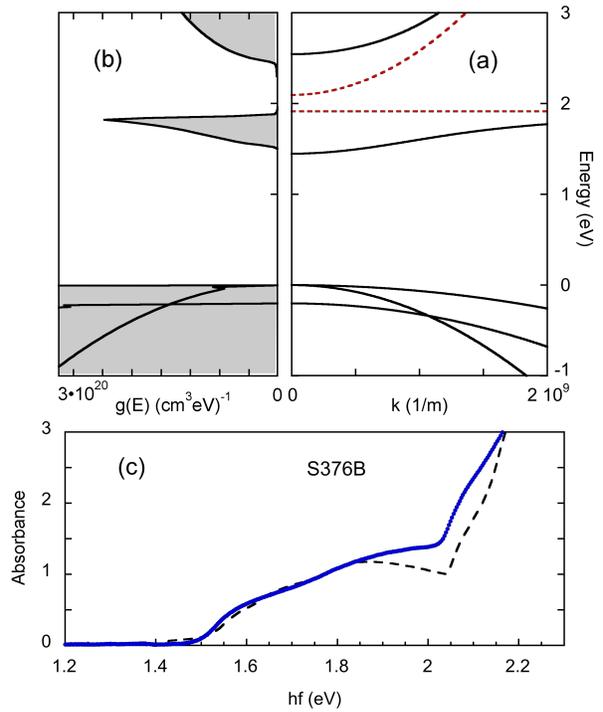

Figure 1. (a) Model energy band diagram for GaP$_y$As$_{1-x-y}$N$_x$ sample S205B (x=1.2%, y=33%) from the Band Anti-Crossing Model (solid lines). Dashed lines show the unperturbed conduction band and N-defect level. (b) Density of states versus energy. (c) Optical absorbance versus energy. The solid line is the measured absorbance, and the dashed line is the model absorbance of the sample and the substrate.

Figure 2. (a) Model energy band diagram for GaP$_y$As$_{1-x-y}$N$_x$ sample S376B (x=3.6%, y=49%) from the Band Anti-Crossing Model (solid lines). Dashed lines show the unperturbed conduction band and N-defect level. (b) Density of states versus energy. (c) Optical absorbance versus energy. The solid line is the measured absorbance, and the dashed line is the model absorbance of the sample and the substrate.

## 2.    Theoretical Model

The dilute III-V$_{1-x}$ – N$_x$ alloy semiconductors studied here have been described in terms of a two-level band anticrossing model (BAC) [1,2,21]. The BAC model can be derived from the multi-impurity Anderson model in the coherent potential approximation. In this model, the restructuring of the conduction band results from the interaction between localized N orbitals and extended conduction-band states of the semiconductor host. The resulting energy bands can be found from perturbation theory:

$$E_k^{\pm} = \frac{1}{2}\left[ E_k^c + E_N \pm \sqrt{(E_k^c - E_N)^2 + 4V^2 x} \right] \; , \qquad (1)$$

where $E_k^C$ is the unperturbed band energy, $E_N$ is the energy of the nitrogen defect state, $V$ is the interaction strength, and $x$ is the nitrogen fraction. The main effect of the interaction is to split the conduction band into $E_+$ and $E_-$ bands separated by a gap. Calculated energy bands, density of states functions and optical absorption for the samples used in this study are shown in Figures 1 and 2.

| Sample | $x$ | $y$ | $V$ (eV) | $E_N$ (eV) | $\beta\rho_0(E_N)$ (eV$^{-1}$) |
|--------|------|------|----------|------------|----------------------------------|
| S205B  | 1.2% | 33%  | 2.85     | 1.988      | 10$^{-3}$                        |
| S376B  | 3.6% | 49%  | 2.85     | 1.915      | 10$^{-3}$                        |

**TABLE 1.** BAC model parameters used for the two samples investigated in this work. Energies are relative to the Valance Band maximum.

In the Anderson model, conduction band states acquire a homogeneous linewidth due to the mixing between the localized defect states and the extended band states. The density of states in the restructured conduction bands $\rho\left(E_k^\pm\right)$ has been calculated by Wu using Greens function techniques [2]:

$$\rho(E_k^\pm) = \frac{1}{\pi} \int_{band} \rho_0(E_{k'}^c) \cdot \text{Im}[G(E_k^\pm, E_{k'}^c)] \, dE_{k'}^c \ , \qquad (2)$$

where $\text{Im}[G(E_k^\pm, E_{k'}^c)]$ is the imaginary part of the Greens function

$$G_{kk}(E, E_k^c) = \left[ E_k^\pm - E_k^c - \frac{V^2 x}{E_k^\pm - E_N - i\pi V^2 \beta \rho_0(E_N)} \right]^{-1} . \qquad (3)$$

Here $\rho_0(E_k^c)$ is the density of states in the unperturbed conduction band, $\rho_0(E_N)$ is the unperturbed density of states at the nitrogen defect state energy, and $\beta$ is a constant of order 1. In our samples the defect energy $E_N$ lies close to or below the unperturbed conduction band edge, so that $\rho_0(E_N) << 1$ (1/unit-cell·eV). We found the predicted optical absorption coefficient to be insensitive to $\rho_0(E_N)$ for small values of this parameter and chose $\beta\rho_0(E_N) = 10^{-3}$ in our model. The valance-band density of states was calculated in spherical parabolic band approximation.

Our samples consist of a GaPAsN layer grown on a series of GaPAs layers on a GaP substrate. The optical absorbance was computed as the sum of the absorbance in each layer of the sample:

$$a(E_J) = \alpha_N(E_J)d_N + \sum_i \alpha_{GaP_{(1-y_i)}As_{y_i}}(E_J)d_i + \alpha_{GaP}(E_J)d_{substrate}. \qquad (4)$$

The first term on the right hand side represents the nitrogen-doped layer; the second term is a sum over the GaPAs layers. The last term is the substrate absorption. Absorption in the GaPAs layers was assumed to be proportional to the joint density of states calculated in the effective mass approximation, with the proportionality constant determined by fitting the near band-edge absorption in GaAs. Absorption in the substrate was modeled using published optical constants for GaP.[22]

The optical absorbance in the GaPAsN layer was calculated using the BAC model following the procedure described by Wu[2]. Although the nitrogen defect states are localized and randomly distributed, $k$-conservation is restored in the coherent potential approximation. The absorption coefficient is:

$$\alpha_N(E_J) = \frac{C}{E_J} \sum_v \sum_{+,-} \rho_J(E_J^{\pm}) |M|^2 (1 - f_k^{v,h} - f_k^{\pm,e}), \qquad (5)$$

where $C$ is a constant, $f_k^{v,h}$ is the hole occupancy of state $k$ in valance band $v$ and $f_k^{\pm,e}$ is the electron occupancy of state $k$ in the $E_+$ or $E_-$ band. The joint density of states for the transition was calculated to be

$$\rho_J(E_J^{\pm}) = \int_{band} \rho(E_k^{\pm}) \, \delta(E_J - E_k^{\pm} + E_k^v) \, dE_k^c . \qquad (6)$$

The matrix elements for the transition are determined by the localized and extended components of the electron wavefunctions. Perturbation theory yields:

$$M_{v,E_+} = \cos(\theta/2)M_{v,c} + \sin(\theta/2)M_{v,N} ,$$
$$M_{v,E_-} = -\sin(\theta/2)M_{v,c} + \cos(\theta/2)M_{v,N} , \qquad (7)$$
$$M_{E_+,E_-} = \left[ \cos^2(\theta/2) - \sin^2(\theta/2) \right]M_{c,N}$$

where $\tan(\theta) = \dfrac{2V}{E_k^c - E_N}$, $M_{v,c}$ is the optical matrix element between valance band $v$ and the unperturbed conduction band, and $M_{v,N}$ and $M_{c,N}$ are matrix elements between the band states and the nitrogen defect state. Figures 1 and 2 show good agreement between the model and experiment by treating the nitrogen fraction of each sample as an adjustable parameter, and assuming $M_{v,N} << M_{v,c}$.

Induced absorption between $E_-$ and $E_+$ should be observable after optical population of the intermediate band. For optical transitions that conserve energy and momentum, we should find

$$\alpha_N(E_J) = \frac{D}{E_J} \rho_J(E_J^\pm) \left| M_{E+,E-} \right|^2 (f_k^{-,e} - f_k^{+,e}) \, , \qquad (8)$$

where $D$ is a constant.  As discussed below, we observe induced absorption in one sample that cannot be fit by equation (8), suggesting it arises from intermediate to conduction band transitions that do not conserve $k$.  We have modeled this as a spectrally featureless absorption process, $\alpha \propto n_- / E_J^2$.

## 2. Experiment

Epitaxial $GaP_yAs_{1-x-y}N_x$ layers were grown in a Varian Gen-II molecular beam epitaxy system modified to handle gas phase $AsH_3$ and $PH_3$[19].  Starting at the GaP substrate, an 0.3-µm-thick GaP buffer layer was grown at 580°C, followed by a 1.5-µm-thick linearly graded $GaP_yAs_{1-y}$ layer in which $y$ increased from zero to the final composition, and then by an 0.5-µm thick $GaP_yAs_{1-y}$ layer grown at 520°C. The 0.5-µm-$GaP_yAs_{1-x-y}N_x$ active layer was then grown at 520 °C using RF-plasma activated N.  The sample structure was confirmed by Rutherford Back Scattering (RBS), which determined the concentrations of Ga, As and P as a function of depth. For S205B the N concentration in a sample from the same wafer was measured to be $x = 1.5\%$ by nuclear reaction analysis (NRA) using the $^{14}N(\alpha,p)^{17}O$ reaction with a 3.72MeV $^4He^2$ beam.  This result is close to the value ($x = 1.2\%$) obtained by fitting the optical absorption in our sample.

Samples for optical measurements were mounted epi-side down on quartz slides with optical epoxy and the substrates were thinned to ~20 µm.  Transmission and reflectance measurements were performed with a Perkin-Elmer 950 UV/VIS/NIR spectrophotometer to determine the absorbance over the spectral range 1.0eV-2.5eV. The measured absorbance of each sample was compared to the simulated absorbance in a model structure (Figures 1c, 2c).

Optical pump/optical probe transient absorption measurements of these samples were performed using a Helios transient absorption system (Ultrafast Systems). In this system a 1-KHz repetition rate amplified femtosecond Ti-Sapphire laser (Coherent Evolution) creates ~5mJ, 100fs pulses centered at $\lambda$~800nm. Pulses are divided by a beamsplitter to generate a tunable pump pulse in an optical parametric amplifier, and a super-continuum white-light probe pulse.  The white-light pulse is further divided into a sample pulse and a reference pulse. The pump pulse propagates through a delay stage and uniformly illuminates the region of the sample interrogated by the sample pulse. Pump-pulse photon energies of 1.91eV (650nm) and 2.82eV (440nm) were used, and measurements were recorded at pump intensities ranging from ~0.5 to 5mJ/cm². Sample and reference pulses are recorded by separate CCD spectrographs allowing transmission spectra to be obtained for each laser pulse. An optical chopper blocks alternate pump pulses to determine the change in the transmission spectrum of the sample due to the pump as a function of the delay between the pump and probe pulses. Transient absorption spectra were recorded in the visible spectral range (1.5-2.5eV) and the near-infrared spectral range (0.8-1.5eV) in separate experiments using independent systems.

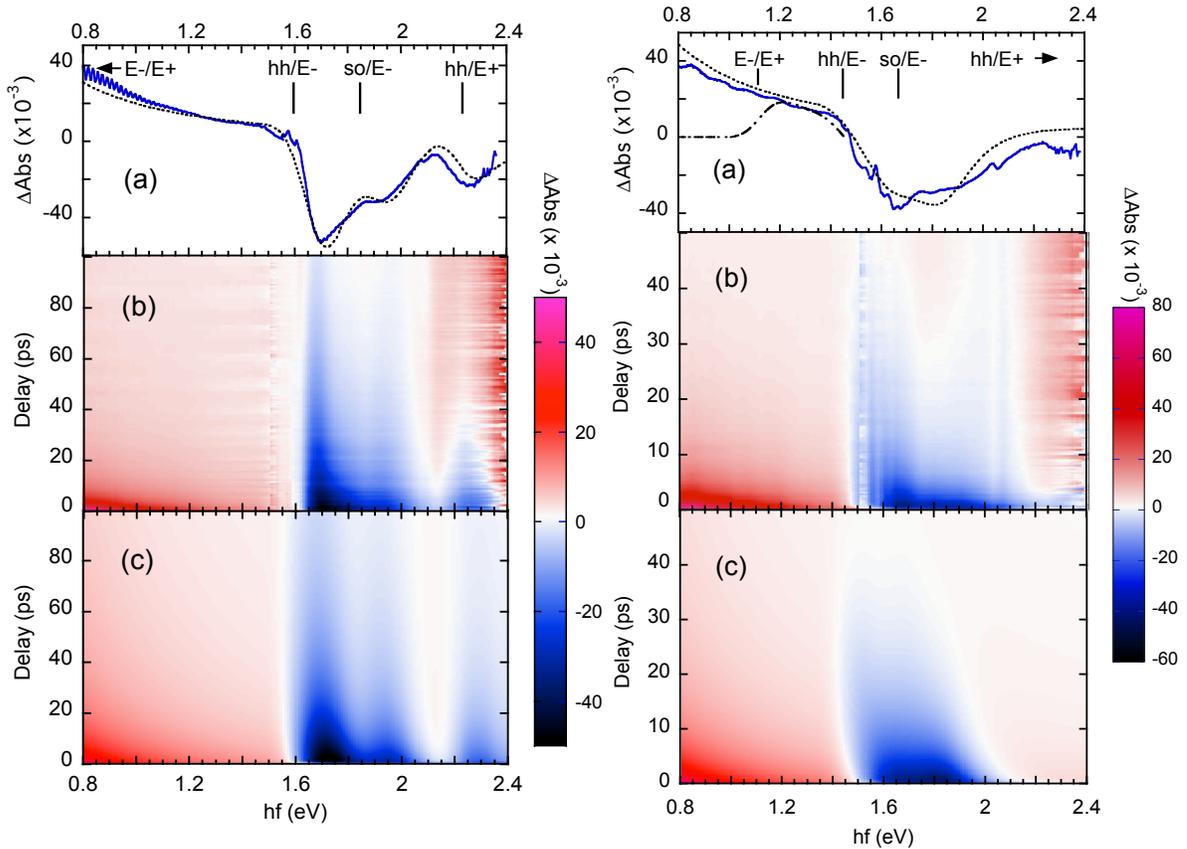

Figure 3. Transient absorbance of sample S205B. (a) Absorbance change 2ps after optical excitation. Solid line is the measurement, while the dashed line is our model. Markers show onset energies for interband transitions. The $E_-$-$E_+$ onset is outside the measured range. (b) Measured absorbance change for pump-probe delays of 0 – 100ps. (c) Model prediction.

Figure 4. Transient absorbance of sample S376B. (a) Absorbance change 2ps after optical excitation. Solid line is the measurement, dotted line is our model, dashed line is a model including only $k$-conserving $E_-/E_+$ transitions. Markers show onset energies for interband transitions. The $hh$-$E_+$ transition is outside the measured range. (b) Measured absorbance change for delays of 0 – 50ps. (c) Model prediction.

## 3. Results:

Figure 3 summarizes the transient absorption data for sample S205B (GaP$_{0.33}$As$_{0.66}$N$_{0.012}$). In this measurement the pump photon energy was 2.82eV and the pump intensity was ~1 mJ/cm$^2$. The spectra show an optically induced *decrease* in absorption at probe photon energies $hf > 1.6\text{eV}$ and an *increase* in absorption at energies $hf < 1.5\text{eV}$. The absorption saturation observed at $hf > 1.6\text{eV}$ contains two distinct bands with maxima at 1.7eV and 2.25eV. A third, partially resolved band is observed at 1.9eV. The induced absorption at $hf < 1.5\text{eV}$ is stronger at low frequencies but shows no other spectral features. Figure 4 summarizes the transient absorption data for sample S376B ($GaP_{0.49}As_{0.474}N_{0.036}$). The results are qualitatively similar, with an optically induced *decrease* in absorption at probe photon energies $hf > 1.5\text{eV}$ and an *increase* in absorption at energies $hf < 1.5\text{eV}$. We do not observe a distinct higher energy band in the transient absorption spectrum of S376B below $hf = 2.4\text{eV}$, the limit set by substrate absorption.

Our measurements also probe the dynamics of the transient absorption signal. For pump excitation at $hf=2.82\text{eV}$ (440nm) in sample S205B (see Figure 3b) the absorption change builds up over ~1ps following photoexcitation and decays on a timescale of 10 – 100ps. The spectrum of the transient absorption decrease at $hf > 1.6\text{eV}$ narrows toward lower frequency with increasing delay. For pump excitation at $hf = 1.91\text{eV}$ (650nm, not shown) the absorption change builds up faster, reaching a maximum only ~0.3ps after the pump pulse, and the transient absorption signal is concentrated at lower frequencies. Similar trends are observed in the transient absorption spectra of sample S376B ($GaP_{0.49}As_{0.47}N_{0.036}$) (Figure 4b).

## 4. Discussion:

We modeled the transient absorption by fitting valance, intermediate and conduction band populations and a single carrier temperature $T_e = T_h$ at each moment in time. Carrier populations in optically excited III-V semiconductors thermalize in tens of femtoseconds [23], much faster than the time scale investigated here. We assumed thermalized (Fermi-Dirac) carrier distributions in each band, found the distribution functions, and calculated the optical absorption using equation (4). In addition, because the strength of the $E_-$-$E_+$ absorption depends on a matrix element $M_{cN}$ not determined by our equilibrium measurements, the ratio of the induced absorption strength to the intermediate band population was used as an adjustable parameter for each sample.

The model fits the transient absorption data for sample S205B well (Figure 3a). Saturated absorption features are observed just above the onset energies predicted for the heavy-hole to $E_-$ transition ($hh$-$E_-$), the split-off band to $E_-$ transition ($so$-$E_-$) and the heavy-hole to $E_+$ transition ($hh$-$E_+$). The induced absorption observed at $hf < 1.5\text{eV}$ is consistent with the $E_-$-$E_+$ transition. For sample S376B (Figure 4a), the model predicts the observed decrease in absorption at $hf = 1.4$-$2\text{eV}$ due to saturation of the $VB$-$E_-$ transition, although model spectra are smoother than the experimental data. It also predicts saturation of the $VB$-$E_+$ transition at $hf = 2.5\text{eV}$, but this is inaccessible due to substrate absorption. Induced absorption between the $E_-$-$E_+$ bands is predicted at $hf = 1.1$-$1.5\text{eV}$. However, the experimental induced absorption does not cut-off below 1.1eV, as predicted for transitions that conserve $k$. Instead, we are able to fit the induced absorption in both samples by assuming $\alpha \propto n_- / E_j^2$, indicating that the induced absorption is proportional to population of the intermediate band.

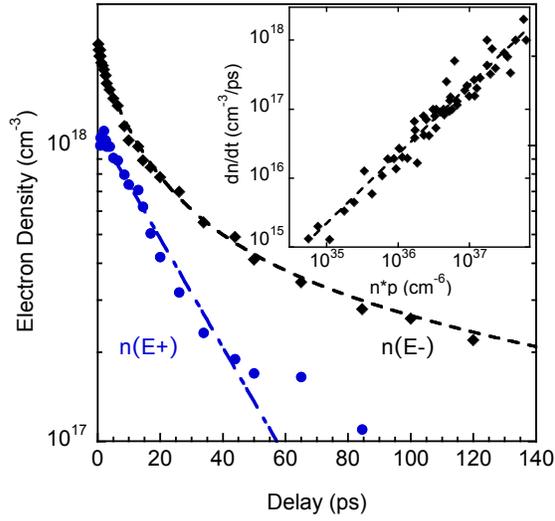

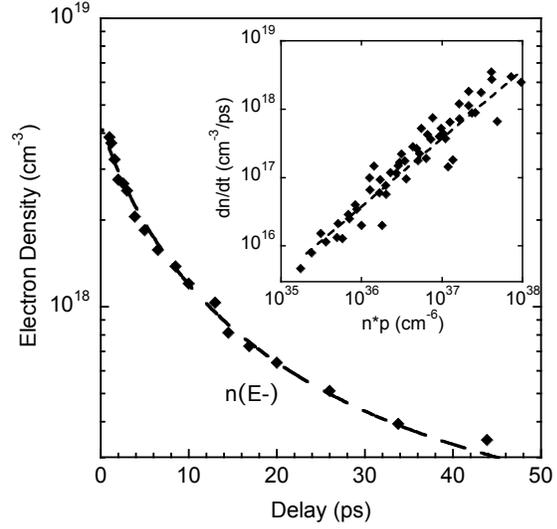

Figure 5. Carrier population versus time for sample S205B from model fits to the transient absorption versus delay at pump intensity of $1mJ/cm^2$. Black diamonds: intermediate band population n(E-). Black dashed line is a bimolecular recombination model. Blue circles: conduction band population n(E+). Blue dashed line is an exponential decay model. INSET: rate of change of n(E-) versus n(E-)·p. Points are obtained from measured carrier populations for pump intensities 0.5 – 5mJ/cm². Linear fit (dashed line) is the bimolecular recombination model.

Figure 6. Carrier population versus time for sample S376B measured at pump intensity of $1mJ/cm^2$. Black diamonds: intermediate band population n(E-) versus delay. Black dashed line is a bimolecular recombination model. INSET: rate of change of n(E-) versus n(E-)·p. Points are obtained from measured carrier populations for pump intensities 0.5 – 5mJ/cm². Linear fit (dashed line) is the bimolecular recombination model.

The transient absorption evolves due to carrier redistribution within the bands and carrier recombination. Model fits match the evolution of the transient absorption well for S205B (Figure 3c). For S376B the model fit is spectrally smoother than the experimental data (Figure 4c). We used these fits to extract electron concentrations in the $E_+$ and $E_-$ bands and the carrier temperature at each delay value for each sample. The intermediate and conduction band populations in sample S205B extracted from these fits are plotted in figure 5. For pump irradiance $I \leq 1mJ/cm^2$, the conduction band population versus time is found to decay exponentially with time with $n(E_+) \propto e^{-t/\tau_{CB}}$ and time constant $\tau_{CB} = 23ps$. At higher pump irradiance the conduction band population decays faster and the decay is non-exponential. In contrast, the electron population in the intermediate band does not decay exponentially, but instead fits a bimolecular recombination process $dn(E_-)/dt = -r\left(n(E_-) \cdot p\right)$, where $r$ is the recombination constant and $p = n(E_-) + n(E_+)$. A plot of $dn(E_-)/dt\ t$ versus $n(E_-) \cdot p$ for the full range of pump intensities investigated (inset, figure 5) yields $r = 2.3 \cdot 10^{-8}$ cm$^{-3}$/s. Figure 6 shows the intermediate band population versus time for sample S376B. The intermediate band dynamics also fit a bimolecular recombination process, although the value of the recombination constant depends on our model for the (unobserved) conduction band population. Assuming a similar exponential decay of the conduction band population in S376B as in S205B yields $r = 3.5 \cdot 10^{-8}$ cm$^{-3}$/s. Bimolecular recombination in semiconductors is often associated with radiative recombination, but radiative $r$-values are typically ~100x smaller than those observed here. We estimate $r_{radiative} = 3 \cdot 10^{-10}$ cm$^{-3}$/s for S205B using the van Roosbroeck-Shockley relation [24]. We suggest that trap-mediated recombination is responsible for the higher recombination rates observed in our samples.

The ~1ps buildup time of the transient absorption signal arises from hot carrier effects. The carrier temperature following photoexcitation versus delay for sample S205B is shown in Figure 7. We find a carrier temperature of 2160K at 0.2ps after photoexcitation. The carriers initially cool rapidly, reaching 500K at 1ps and 400K at 2.6ps. Our data can be described well by a cooling curve calculated for hot carriers in GaAs by J. Shah, *et. al.*[23] (see Figure 7), where we have shifted the origin of the time axis for the calculated cooling curve by 0.3ps to match our higher initial temperature. In GaAs hot carriers rapidly cool by optical phonon emission until the carrier kinetic energies drop below the optical phonon energy. Carrier cooling in our GaPAsN samples appears to follow the same process.

As noted above, for sample S376B the induced absorption does not cut off below the energy of the smallest *k*-conserving interband transition predicted by our model. This leads us to consider two other mechanisms for the induced absorption: free-carrier absorption by the photoexcited holes, and non-*k* conserving $E_- $-$E_+$ transitions by the electrons. Note that intraband absorption by free *electrons* cannot be responsible for the induced absorption because the transition energies exceed the width of the intermediate band. It is clear that intraband (Drude) and intervalence band transitions by optically excited holes will produce absorption. These processes have been studied in *p*-type bulk III-V semiconductors [25] [26], and they produce weak optical absorption with $\alpha \propto n/E^2$ in the spectral region $hf$ =0.8-1.5eV.  However, the induced absorption in S205B and S376B is too strong to be explained by this mechanism. Intervalance band absorption varies with hole density, while Drude absorption depends on both carrier density and momentum scattering time. For hole densities predicted by our model no scattering time produces free hole absorption larger than ~10% of the observed induced absorption. In contrast, weaker induced absorption observed in our reference samples (GaPAs samples with no nitrogen doping, not shown) in the IR spectral range is consistent with free carrier absorption.

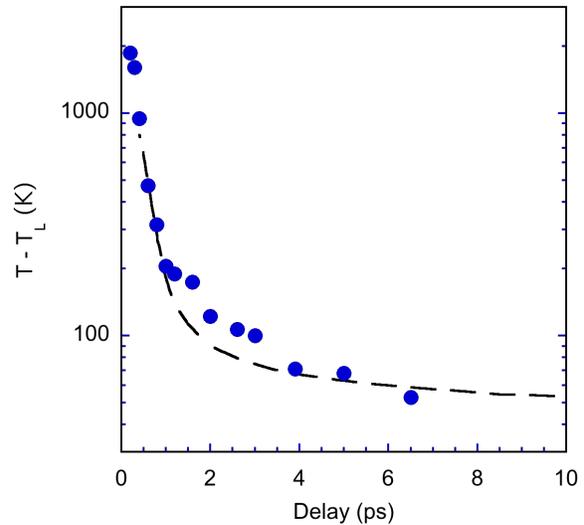

Figure 7. Temperature difference between the electron gas and the lattice versus time after optical excitation for sample S205B measured at pump intensity of 1mJ/cm$^2$. Points were obtained from model fits to the transient absorbance. Black dashed line is a calculated cooling curve for electrons in GaAs [J. Shah, Ultrafast Spectroscopy of Semiconductors and Semiconductor Nanostructures, 2 ed. (Springer, Berlin, 1999)].

Alternately, we suggest that non $k$-conserving interband transitions of electrons between $E_-$ and $E_+$ contribute to the optical absorption. The energy range of all possible interband transitions extends down to the $E_-$ -$E_+$ bandgap, which is below the experimentally accessible spectral range for both samples.  We speculate that disorder in our samples may relax $k$-conservation for $E_-$ -$E_+$ optical transitions.   Transitions near the $E_-$ -$E_+$ bandgap also involve initial and final states that have the highest contribution from the localized nitrogen orbitals, and thus most strongly violate the coherent potential approximation.  We conclude that our data are consistent with this mechanism.

Finally, we consider the impact of these results on the feasibility of a GaPAsN intermediate band solar cell.  In order to efficiently collect photoelectrons, the photocarrier lifetime in the conduction band should be larger than the time required for a carrier to cross the active layer of a device, of order ~100ps, longer than the conduction band lifetimes measured in our samples. In addition, an efficient IBSC requires photo-carriers to be stored in the intermediate band until excitation to the conduction band occurs.  In the IBSC concept, the materials are doped $n$-type to place the Fermi level in the middle of the intermediate band, so that $n \gg p$ under normal illumination.  If the bimolecular recombination process we observe is dominant under these conditions, the photocarrier lifetime in the intermediate band will be $\tau \sim 1 / n \cdot r$ . Thus, the electron concentration must be limited to $n < 5 \cdot 10^{17} \text{cm}^{-3}$ to give a photocarrier lifetime longer than 100ps.  This is much lower than the doping level required to place the Fermi level in the middle of the intermediate band ($\sim 10^{19} \text{cm}^{-3}$).  We conclude that the minority carrier lifetime in these samples would also be too short for an efficient solar cell.  However, since the carrier recombination is most likely controlled by defects rather than by intrinsic processes it may be possible to achieve longer minority carrier lifetimes by optimizing the materials synthesis.   In addition, device structures with composition and doping gradients can be used to improve charge collection.

## 5.      Conclusions:

We have used transient absorption spectroscopy to probe carrier lifetimes in two samples of the intermediate band semiconductor $GaP_yAs_{1-x-y}N_x$.  An optical pump pulse excited electrons from the valance band to the intermediate and conduction bands, and we recorded absorption saturation of the $VB$-$E_-$ and $VB$-$E_+$ transitions and induced absorption that we associate with the $E_-$-$E_+$ transition. We modeled the optical properties of our samples using the band anti-crossing model to extract carrier densities as a function of time.      Following photoexcitation of GaP$_{0.32}$As$_{0.67}$N$_{0.012}$ (S205B) with <1mJ/cm$^2$ pulses, we find that the electron population in the conduction band decays exponentially with time constant $\tau_{CB}$ =23ps. The electron population in the intermediate band exhibits bimolecular recombination with recombination constant $r = 2 \cdot 10^{-8}$ cm$^{-3}$/s. Measurements of GaP$_{0.54}$As$_{0.46}$N$_{0.036}$ (S376B) found that the intermediate band electron population exhibits bimolecular recombination with recombination constant $r = 3.5 \cdot 10^{-8}$ cm$^{-3}$/s. These rapid recombination processes would strongly limit the efficiency of an intermediate band semiconductor made from this material.

Work was performed in the Electronic Materials Program and the Molecular Foundry at Lawrence Berkeley National Laboratory and was supported by the Office of Science, Office of Basic Energy Sciences, of the U.S. Department of Energy under Contract No. DE-AC02-05CH11231. KMY acknowledges the support of the General Research Fund of the Research Grants Council of Hong Kong SAR, China, under project number CityU 11303715. The samples were synthesized in Prof. Tu's laboratory at UC San Diego and the experiments were carried out by J. H. and A. S. The interpretation of the experimental results was done by J. H. and W. W. The other authors contributed to the material characterization and discussed the results. The authors acknowledge the support of the Joint Center for Artificial Photosynthesis at U.C. Berkeley for transient absorption measurements using an infrared probe.